\documentclass{elsart}
\usepackage{epsfig}
\usepackage{amssymb}
\gdef\bc{\begin{center}}
\gdef\ec{\end{center}}
\gdef\be{\begin{eqnarray}}
\gdef\ee{\end{eqnarray}}
\def\slrt{spin-lattice relaxation time}
\def\t1{$T_1$}
\def\mag{magnetization}
\def\IR{inversion recovery}

\begin{document}
\begin{frontmatter}
\title{A fast method for the measurement of long spin-lattice relaxation times by Single Scan Inversion Recovery experiment}
\author[physics]{Rangeet Bhattacharyya}
\author[physics,sif]{Anil Kumar}
\address[physics]{Department of Physics, Indian Institute of Science, Bangalore - 560012, India}
\address[sif]{Sophisticated Instrument Facility, Indian Institute of Science, Bangalore - 560012, India}
\begin{abstract}
A new method of measuring long spin-lattice relaxation times ($T_1$) is proposed.
Being a single scan technique, the method is at least one order of magnitude faster than the conventional technique. 
This method (Single-Scan or Slice Selected Inversion Recovery or SSIR) relies on the slice selection technique. 
The method is experimentally verified and compared with the time taken by the conventional measurement.  
Furthermore, it is shown that the conventional Inversion Recovery (IR) method which suffers from 
effects of r.f. inhomogeneity can also be improved by measuring the magnetization of only a central slice. 
\end{abstract}
\begin{keyword}
Spin-lattice relaxation time, Inversion recovery experiments, single scan, slice-selection.
\PACS 76.60.-k \sep  
\end{keyword}
\date{}
\end{frontmatter}
\section{Introduction}
Measurement of the \slrt\ is of enormous importance in the field of nuclear magnetic resonance (NMR) spectroscopy. Since the
historical introduction of Bloch equations to describe the time evolution of the perturbed
magnetization, the knowledge of \slrt\ plays a significant role in understanding the dynamics
of spin systems \cite{bloch1}. 

To measure the \slrt, the most commonly employed technique is the \IR\ (IR) experiment, which is a two-pulse experiment \cite{NB}. 
In this 
method, a block of (d$_1$ - 180$^{\circ}$ - $\tau$ - 90$^{\circ}$ - acquisition) is repeated for various $\tau$ values. The 
magnetizations for various $\tau$ values measure the recovery from inverted state towards the equilibrium value. 
The most time consuming step in this experiment is the recovery period $d_1$ during which the magnetization is repeatedly required
to recover to its equilibrium value, necessitating $d_1 \ge 5T_1$.

The measurement of long \slrt{s} therefore takes significant time. Many methods have been suggested which speed-up 
the process, but with limited success. For example, a null measurement in an \IR\ experiment yields approximate value in a short time
\cite{null}.
Saturation recovery methods take less time than \IR\ since the recovery time is replaced by a 
saturation burst which usually takes $\sim 2$ \t1 \cite{NB}. 
Progressive saturation method is the most commonly used method for long \t1\ measurements \cite{NB}. 
Freeman and Wittekeok suggested a fast method for \t1\ measurements, which works for spins which gives extremely narrow lines 
($\sim$ 0.1 Hz) in the spectrum \cite{FW}.
But for accurate 
measurements the \IR\ method is still most commonly employed technique \cite{FB}.

Recently Frydman~{\it et~al} have utilized the echo-planar imaging protocols to speed-up acquiring of two and multi-dimensional NMR 
experiments in which different slices of the sample (separated by using a linear gradient) are excited sequentially to encode indirect 
time evolution information, followed by a series of frequency-discriminated acquisitions, which are done in a single scan 
\cite{fryd1,fryd2}. Experimental times have been reduced from hours to seconds, without significant loss in S/N ratio. At the same time, 
several new experiments have been developed which speed-up multi-dimensional NMR experiments, such as G-matrix Fourier Transform NMR (GFT)
 and a single scan measurement of diffusion coefficients  \cite{gft,dosy}. 

Following Frydman's idea, we describe here a {\it fast} method of measurement of \slrt, by a single scan 
experiment for samples having sufficient
S/N ratio. The method incorporates the slice selection protocol in a standard \IR\ experiment. Like a conventional \IR\ experiment, a 
$\pi$ pulse is applied to invert the \mag. As the \mag\ is
relaxing back toward the equilibrium value, we observe different slices of the sample (thickness is typically $1/100{\rm th}$ of the 
sample length) at different times.  A selection of 10-20 slices allow us to monitor the journey of \mag\ from 
an inverted state toward equilibrium, and to accurately calculate the \t1\ value.

\section{The method}
\begin{figure}[htb]
\centerline{\epsfig{file=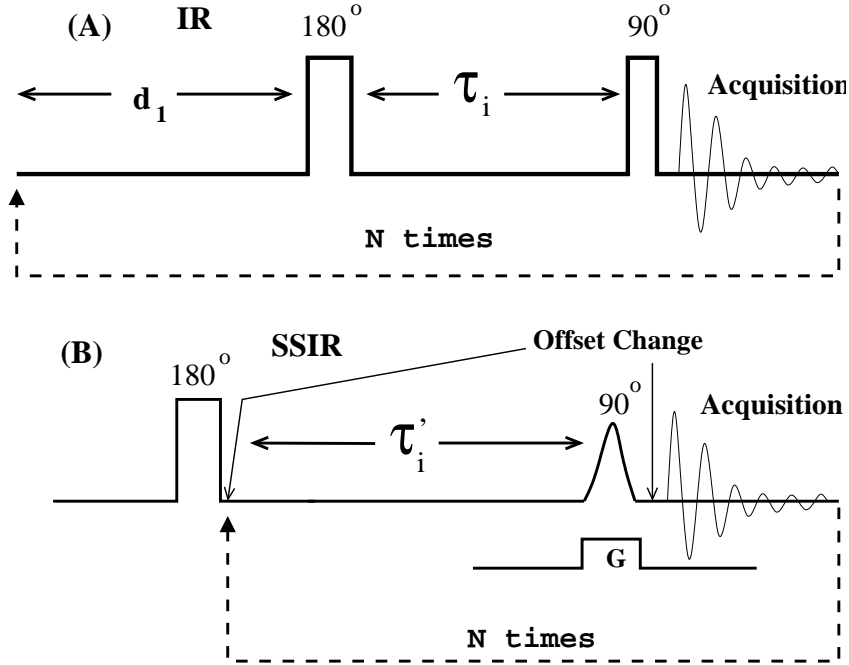,width=0.8\linewidth}}
\caption{(A) Standard Inversion Recovery experiment. $d_1$ is the recycle delay ($\sim$ 5\t1), $\tau_i$ is the variable delay. 
(B) SSIR.  The first $180^{\circ}$ pulse inverts the magnetization of the whole the sample. After a variable delay a thin slice of the 
sample is selected for observation by using a $90^{\circ}$ soft pulse in presence of a gradient(G). The position of the selected slice 
is determined by the offset of the $90^{\circ}$ pulse. The magnetization of the selected slice is observed in absence of a gradient. 
The offset is changed prior to observation.  $N$ slices are selected from the sample to monitor recovery of magnetization from the 
inverted to equilibrium state. Time delay for the $i$th slice is $\tau_i = \tau_{i-1} + \tau^{\prime}_i + T_{aq}$, where $T_{aq}$
is the acquisition time.}
\label{twomethods}
\end{figure}

The basic framework of SSIR is described in Fig. \ref{twomethods}B. The figure also shows the standard \IR\ experiment 
(Fig. \ref{twomethods}A), so that a
direct comparison can be made. The total time delay for $i$th slice in SSIR, has been made equal to the $i$th time delay ($\tau_i$) of the
\IR\ experiment, such that a comparison between the experimental time taken by these two methods can be made. In SSIR 
(Fig. \ref{twomethods}B) the offset change at the beginning of a cycle is to select a particular slice at a specific frequency. 
This offset is incremented in each cycle to select slices. However, just before the acquisition, the offset is to reverted back to the 
original value for recording the spectrum, in the absence of a gradient.

From Fig. \ref{twomethods}, it is apparent that the total experimental time for standard inversion recovery experiment, $T_{IR}$, is 
given by,
\be
T_{IR} = N(d_1 + T_{aq}) + \sum_{i=1}^N\tau_i,
\ee
where $T_{aq}$ is the acquisition time. The time delay for $i$th scan is $\tau_i$ for \IR\ experiment. 
In case of SSIR, the time delay for $i$th slice is given by,
\be
\tau_i = \left(\sum_{j=1}^{j=i}\tau_j^{\prime}\right) + (i-1)T_{aq}.
\ee
It should be noted that in contrast to \IR\ experiment, the calculation of total experimental time for SSIR, does not involve
a summation over all $\tau_i\,$s. 
Hence the total experimental time for SSIR, $T_{SSIR}$, is,
\be
T_{SSIR} =  T_{aq} + \tau_{\scriptscriptstyle N},
\ee
where, $\tau_{\scriptscriptstyle N}$ is the time delay for Nth or final slice.
Taking $\tau_{\scriptscriptstyle N} = 3T_1$,
$d_1 = 5T_1$, 
and 
for the \IR\ experiment assuming an average $\tau_i = T_1$, one obtains,  
\be
&&T_{IR} = N(5T_1+T_{aq} + T_1),\\\cr
&&T_{SSIR} = T_{aq} + 3T_1.
\ee
Assuming $T_{aq} \sim 3T_2^{\star} = aT_1$, where $T_2^{\star} = 1/(\pi\Delta\nu)$, 
$\Delta\nu$ being full-width at half-maximum of the
concerned peak, we get,
\be
T_{SSIR} = \frac{(3+a)}{N(6+a)}T_{IR}
\ee
For $a=1$, $T_{SSIR} = (4/7N)T_{IR}$. However for long \t1's, $a \ll 1$, yielding,
\be
T_{SSIR} = \frac{T_{IR}}{2N}.
\ee
Thus, the SSIR method needs only a fraction of the time needed for the full IR method.

\section{Experimental}
\begin{figure}[htb]
\raisebox{2cm}{\large (A)}
\centerline{\epsfig{file=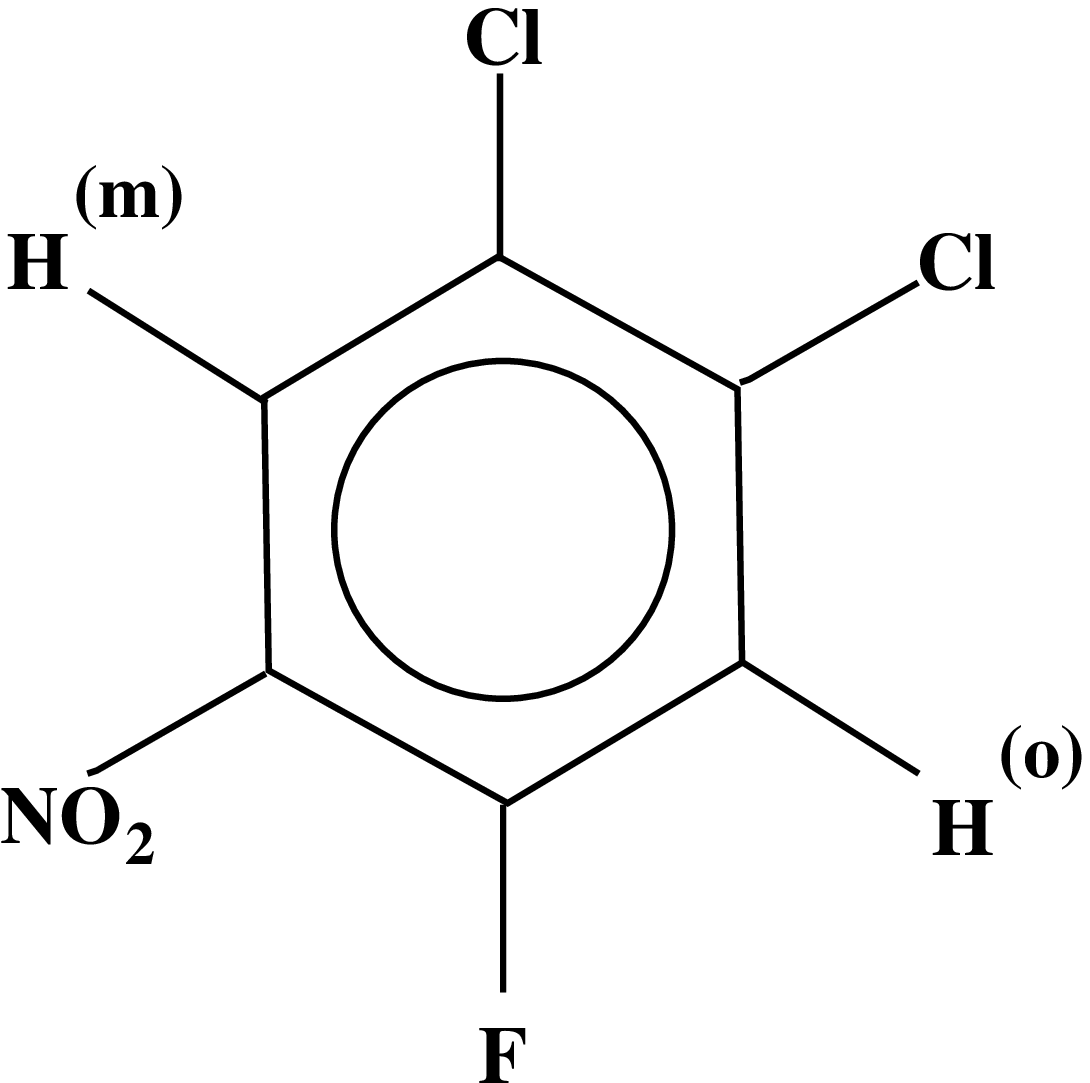,width=0.3\linewidth}}
\vspace*{-3.0mm}
\raisebox{6cm}{\large (B)}
\centerline{\epsfig{file=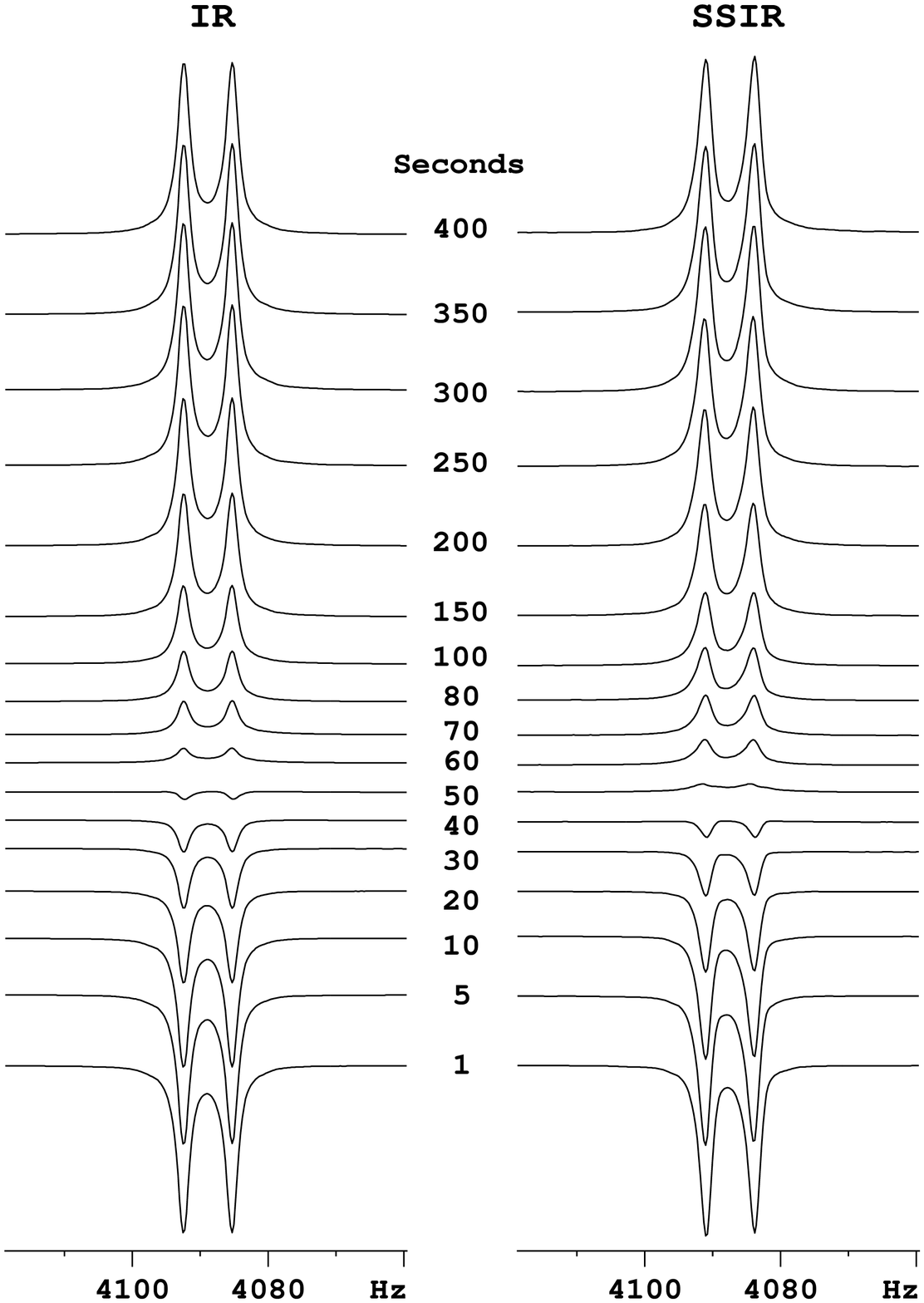,width=0.7\linewidth,height=12cm}}
\caption{(A) The molecule, 4,5-dichloro-2-fluoro-nitrobenzene. (B) Comparison of spectra obtained from \IR\ experiment and spectra 
obtained using SSIR for H$^{(m)}$ proton.
The conventional experimental time in IR was 2 hours 55 minutes, whereas SSIR took only 7.5 minutes. $T_{aq}$ was kept 
1.022 s for both the schemes.  For H$^{(o)}$, the conventional IR experimental time was 1 hour 5 minutes, whereas the SSIR took only 2 minutes.}
\label{stack1}
\end{figure}

Both SSIR and the conventional \IR\ experiment were performed on 4,5-dichloro-2-fluoro-nitrobenzene at 300 K dissolved in CDCl$_3$
and sealed after several freeze pump and thaw cycles. The molecule has three weakly coupled spins (Fig. \ref{stack1}A). 
Spectra of both protons $H^{(\circ)}$ (ortho to fluorine) and $H^{(m)}$ (meta to fluorine), show doublets due to the coupling with
fluorine, the coupling between the protons being small and unresolved.
We have chosen this molecule, since the two protons have long relaxation times ( $\sim$ 25 s and $\sim$ 75 s respectively) at 300 K.
We have earlier observed strong longitudinal cross-correlation effects in this molecule \cite{kavpap}. In the \t1\ measurements 
described below, the
signal of each proton doublet was added up to suppress any multiplet effect due to cross-correlations \cite{anilrev}.

For the SSIR experiment a gradient of 3 G/cm was applied along the z-direction. The effective sample height (the height of the r.f. coil) was
25 mm, yielding the total spectral spread of approximately 30 KHz (at 500 MHz of proton frequency). From the central part of the sample 17
slices of 100 Hz width ($\sim$ 0.08 mm thick) were selected with 500 Hz distance between the slices corresponding to approximately 0.4 mm
distance between the slices. 
To investigate the diffusion effects the SSIR experiment was repeated with various slice distances. Slice distances of
600 Hz, 700 Hz and 1000 Hz were chosen which correspond to 0.47 mm, 0.50 mm and 0.79 mm respectively.
It was assumed that the gradient profile is linear and flat in the region of interest.

The result of standard IR method suffers from the effects of r.f. inhomogeneity, particularly when the sample height is longer than the
r.f. coil height \cite{akjohn}. Methods have been suggested to correct the errors in \t1\ measurements arising due to the r.f. inhomogeneity
effects \cite{akjohn}. Here, we describe a method of obtaining \t1, free from r.f. inhomogeneity effects. In this experiment the
magnetization of the entire sample is inverted using a hard 180$^{\circ}$ pulse in the absence of gradient, and subsequently the
magnetization of only the central slice is detected by using a soft 90$^{\circ}$ pulse in the presence of a gradient. The experiment is
repeated for several values of delay ($\tau$), each after a delay of 5\t1, exactly like in a standard IR experiment. The total experimental
time is thus equal to the standard IR experiment, with the result being free from r.f. inhomogeneity effects. We name this experiment as
IR$^{\star}$.

Fig. \ref{stack1} shows the comparison between the spectra obtained from standard inversion recovery experiment and 
from SSIR for H$^{(m)}$ proton. The H$^{(o)}$ proton spectra (not shown) yield similar results. The \t1\ values 
obtained from these spectra are discussed in the next section.

\section{Results}

\begin{figure}[htb]
\centerline{\epsfig{file=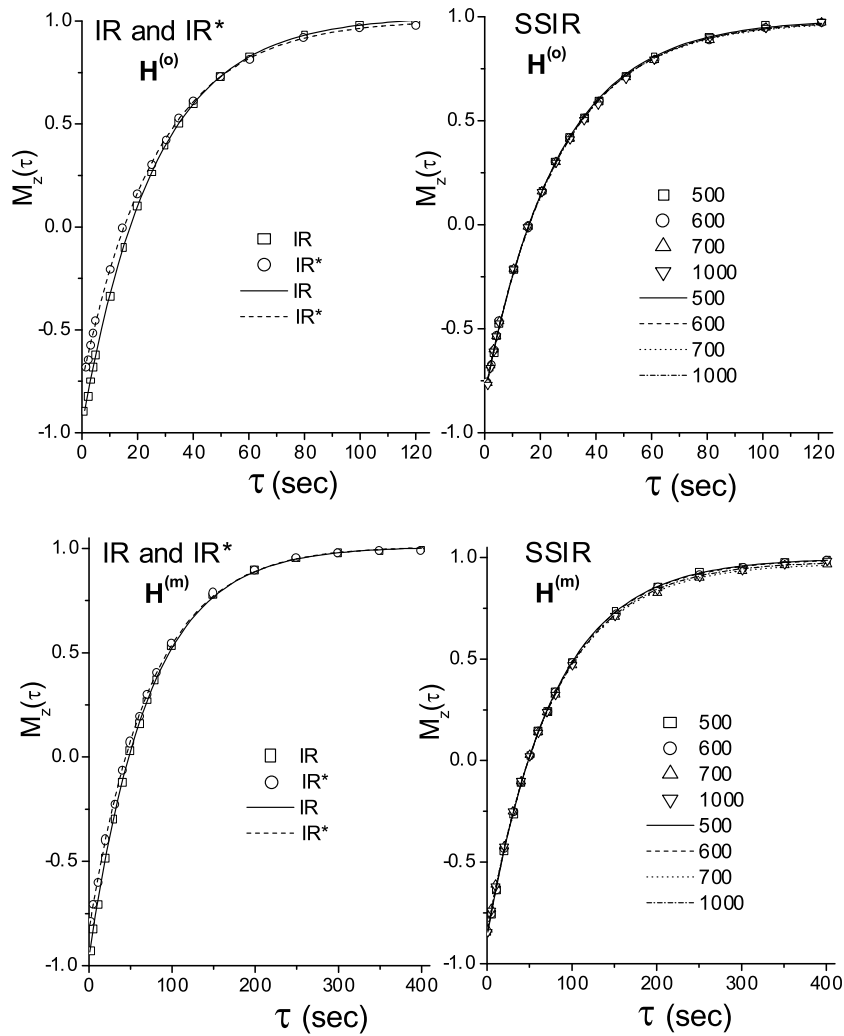,width=0.8\linewidth}}
\caption{Experimental points for the H$^{(o)}$ and H$^{(m)}$ protons for IR, IR$^{\star}$ and SSIR methods and
mono exponential least-square fits for these data are shown in the figure. The values in legends of SSIR plots indicate the 
slice separation in Hz. For the given gradient 500, 600, 700 and 1000 (Hz) correspond to slice separation of 0.39 mm, 0.47 mm, 
0.50 mm and 0.79 mm respectively.}
\label{values}
\end{figure}

The data for the IR, IR$^{\star}$ and SSIR experiments are plotted in Fig. \ref{values} for the H$^{(o)}$ and H$^{(m)}$ protons. 
Results of mono exponential least square fit to these data are summarized in table \ref{comptable}. 

It is seen from table \ref{comptable} that the \t1\ values for all the three methods for the H$^{(o)}$ proton are equal within
experimental errors. For the H$^{(m)}$ proton, however, the IR method gave a slightly lower value. It is well known that the r.f.
inhomogeneity causes errors in IR method and lowers the measured \t1. The IR$^{\star}$ method
is thus superior to IR method. The SSIR method results substantial saving in time without loss of accuracy in \t1\ measurement. 

\section{Discussions}
\begin{figure}[htb]
\centerline{\epsfig{file=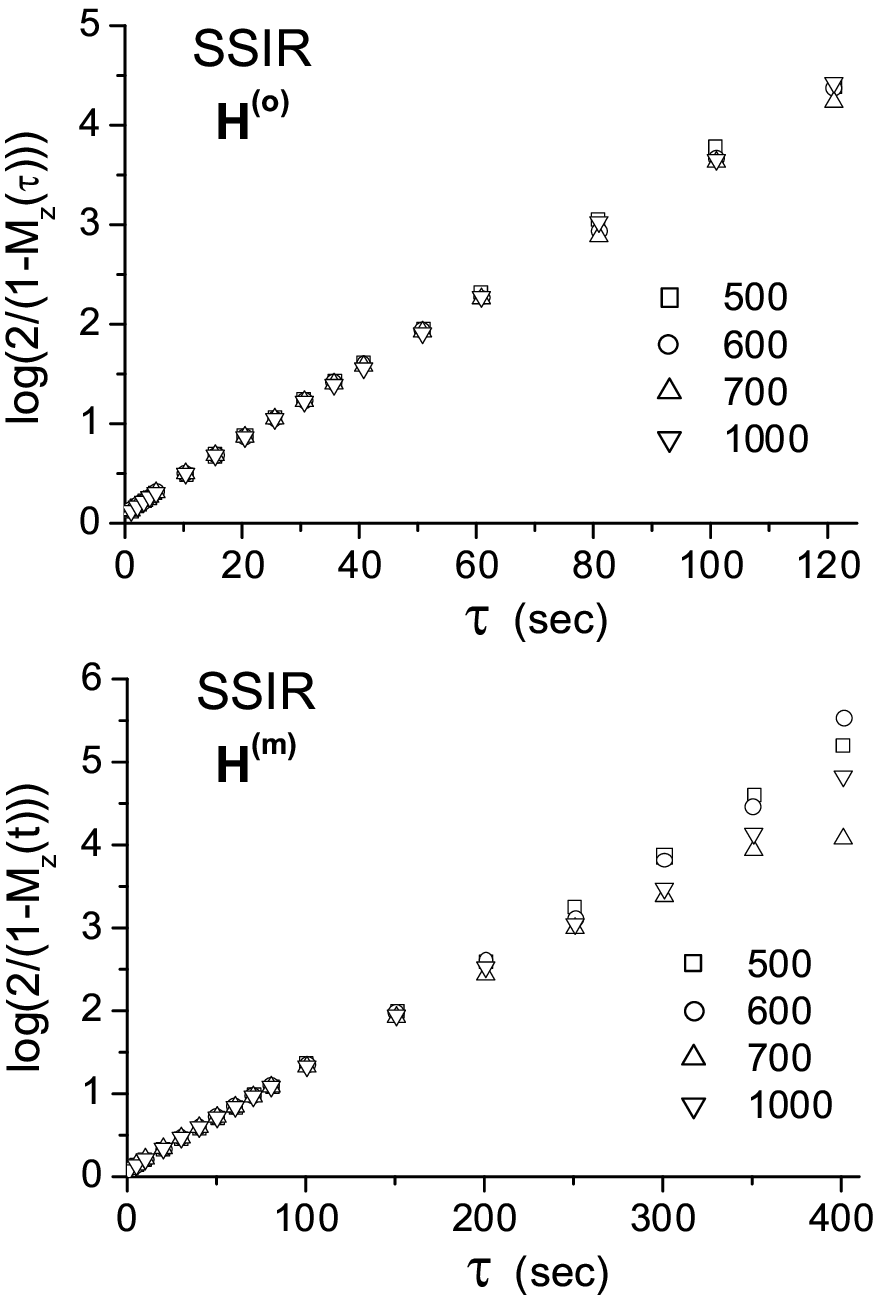,width=0.6\linewidth}}
\caption{Plot of SSIR data in log scale.} 
\label{logplots}
\end{figure}

The SSIR results are also plotted on a logarithmic scale in fig. \ref{logplots}. It is seen that for long values of $\tau$ ($> 200 s$
for H$^{(m)}$) the data show deviation from linearity. It was suspected that this is due to diffusion
of saturation arising from the 90$^{\circ}$ pulses to adjacent slices. To investigate this aspect, four SSIR experiments with
different slice separation (namely, 0.39 mm, 0.47 mm, 0.50 mm and 0.79 mm corresponding to 500, 600, 700 and 1000 Hz) were
carried out. The slice thickness was kept constant to 0.08 mm corresponding to 100 Hz. The results given in table \ref{comptable} show
a slight systematic increase in \t1\ values of the H$^{(m)}$ proton (within experimental errors) with increased slice separation. 
It is thus concluded that the diffusion effects, if any, are small in the present sample.

While the manuscript was under preparation, our attention was drawn to a paper accepted for publication in Journal of Magnetic 
Resonance, by Loening et al, which also describes an experiment similar to our SSIR experiment and reports similar saving in 
experimental time \cite{scoop}. However they have measured smaller \t1\ (\t1\ $\sim$ 2 - 12 s).

\section{Conclusion}
A fast method (Single Scan Inversion Recovery, SSIR in short) for measuring the long \slrt\ has been described. 
Being a single scan experiment, SSIR is order of magnitude faster than the conventional \IR\ method, without any 
significant loss in accuracy.
The method relies on slice selection technique. After an initial
inversion of the total magnetization of the sample, different slices are observed at different times to monitor the complete
relaxation recovery, in a single inversion. The effect of diffusion of saturation from one slice to next have been found to be
small in this sample.

\section{Acknowledgment}
The authors wish to thank Prof K.V. Ramanathan for useful discussions and Dr.P.K. Madhu for drawing our attention to ref. \cite{scoop}.
The use of the 500MHz FTNMR spectrometer
(DRX-500) funded by the Department of Science and Technology, New Delhi, at the Sophisticated Instruments Facility, Indian Institute
of Science, Bangalore is also gratefully acknowledged.

\def\jmr{J. Mag. Res.}
\def\jacs{J. Am. Chem. Soc.}
\def\jcp{J. Chem. Phys.}
\def\pr{Phys. Rev.}

\newpage
\begin{table}[hbt]
\caption{Table of \t1\ values (in seconds) obtained and time taken by various methods (in minutes).}
\label{comptable}
\begin{tabular}{|l|c|c|c|c|}\hline
& \multicolumn{2}{|c|}{H$^{(m)}$}   & \multicolumn{2}{|c|}{H$^{(o)}$} \cr\hline
Experiment & \t1\ (s) & Expt time (min) & \t1\ (s) & Expt time (min) \cr\hline
IR & 75.2 $\pm$ 0.5 & 175 & 26.0 $\pm$ 0.08 & 65 \cr\hline
IR$^{\star}$ & 77.7 $\pm$ 0.6 & 175 & 27.0 $\pm$ 0.1 & 65 \cr\hline
SSIR (500) & 77.4 $\pm$ 0.5 & 7.5 & 26.3 $\pm$ 0.3 & 2 \cr\hline
SSIR (600) & 77.8 $\pm$ 0.5 & 7.5 & 26.4 $\pm$ 0.3 & 2 \cr\hline
SSIR (700) & 77.8 $\pm$ 0.5 & 7.5 & 26.4 $\pm$ 0.3 & 2 \cr\hline
SSIR (1000) & 77.9 $\pm$ 0.5 & 7.5 & 26.6 $\pm$ 0.4 & 2 \cr\hline
\end{tabular}
\end{table}

\newpage

\begin{thebibliography}{00}
\bibitem{bloch1}{F. Bloch}, {\it \pr}, {\bf 70}, 460-474 (1946).
\bibitem{NB}{E. Fukushima and S.B.W. Roeder}, {\it Experimental Pulse NMR, A Nuts and Bolts Approach} (Addison-Wesley, Reading, MA, 1981).
\bibitem{null}{R.K. Harris}, {\it Nuclear Magnetic Resonance Spectroscopy, A Physicochemical view} (Pitman Books Ltd. London, 1983).
\bibitem{FW}{R. Freeman and S. Wittekeok}, {\it \jmr}, {\bf 1}, 238 (1969).
\bibitem{FB}{T.C. Farrar and E.D. Becker}, {\it Pulse and Fourier Transform NMR} (Academic Press, New York and London, 1971).
\bibitem{fryd1}{L. Frydman, T. Scherf and A. Lupulescu}, {\it Proc. Natl.  Acad. Sci. USA}, {\bf 99}, 15858-15862 (2002).
\bibitem{fryd2}{L. Frydman, A. Lupulescu and T. Scherf}, {\it \jacs}, In press.
\bibitem{gft}{S. Kim}, {\it \jacs}, {\bf 125}, 1385-1393 (2003).
\bibitem{dosy}{M.J. Thrippleton, N.M. Loening and J. Keeler}, {\it Magn. Reson. Chem.}, {\bf 41}, 441-447 (2003).
\bibitem{kavpap}{K. Dorai and A. Kumar}, {\it Chem. Phys. Lett.}, {\bf 335}, 176-182 (2001).
\bibitem{anilrev}{A. Kumar, R.C.R. Grace, P.K. Madhu}, {\it Prog. Nuc. Magn. Resno. Spec.}, {\bf 37}, 191-319 (2000).
\bibitem{akjohn}{A. Kumar and C.S. Johnson, Jr.}, {\it \jmr}, {\bf 7}, 55-59 (1972).
\bibitem{scoop}{N.M. Loening, M.J. Thrippleton, J. Keeler and R.G. Griffin}, {\it \jmr}, {Article in press}.
\end{thebibliography}
\end{document}